\documentclass[aps,a4paper,fleqn,usenatbib,pra,twocolumn,nopacs,floatfix,longbibliography]{revtex4-2}
\usepackage{natbib} 
\usepackage{epsfig}
\usepackage{graphicx}
\usepackage{dcolumn}
\usepackage{amsthm,amsmath}
\usepackage{comment}
\usepackage{float}
\usepackage[colorlinks]{hyperref}
\usepackage{xcolor}
\usepackage{ulem}
\usepackage{orcidlink}
\usepackage[justification=raggedright,singlelinecheck=false]{caption}
\usepackage{amsfonts}  
\usepackage{amssymb}  
\usepackage{graphicx}
\usepackage{amsmath,amssymb,amsfonts}
\usepackage{graphicx}
\usepackage{hyperref}
\usepackage{bm}
\begin{document}
\author{Nana Chang\orcidlink{0000-0003-3631-4426}$^1$}
\email{nnchangqq@gmail.com}
\author{Xiaoji Zhou\orcidlink{0000-0001-9175-2854}$^{1,2}$}
\email{xjzhou@pku.edu.cn}
\affiliation{$^{1}$State Key Laboratory of Advanced Optical Communication System and Network, School of Electronics, Peking University, Beijing 100871, China} 
\affiliation{$^2$ Institute of Carbon-based Thin Film Electronics, Peking University, Shanxi, Taiyuan 030012, China}
\title{Magnetic Field Induced Band Deformation in a Lieb Lattice:\\
Aharonov–Bohm Caging and Zeeman Splitting} 
\begin{abstract}
Flat-band systems are highly sensitive to external perturbations, providing a route to study unconventional localization, transport, and spin physics. Lieb lattice, a two-dimensional geometry with an inherent flat band, exemplifies this behavior and is experimentally realizable in ultracold atoms, photonic arrays, and superconducting circuits. In this work, we present a comprehensive study of magnetic-field--induced band deformation in the Lieb lattice by jointly considering orbital Peierls phases and Zeeman spin splitting. A perpendicular magnetic flux generates Aharonov-Bohm caging, confining particles into localized flat-band states at $\phi=\pi$, while Zeeman coupling lifts spin degeneracy and induces spin-resolved energy shifts. The competition between these two mechanisms gives rise to rich band restructuring and tunable spin-selective flat-band phenomena. These results establish the Lieb lattice as a controllable setting for spin-selective transport and magnetic-field engineering in synthetic quantum platforms such as ultracold atoms, photonic lattices, and superconducting circuits, offering guiding principles for quantum simulation and the corresponding experiments, which opens the avenue for controlled engineering of spin-resolved localization and flat-band 
physics in synthetic quantum matter.
\end{abstract}
\date{\today}
\maketitle
\section{Introduction}
Flat-band systems have attracted broad interest in condensed matter physics due to their quenched kinetic energy amplifies interaction effects, enabling strongly correlated phases, geometric frustration, and unconventional transport properties\cite{chen2023decoding, neupert2011fractional, chang2021nonlinear, leykam2018artificial, leykam2013flat, flach2014detangling, mallick2022antipt, mallick2021wannierstark}, where the vanishing kinetic energy enhances interaction effects, making them fertile ground for realizing exotic quantum states. A paradigmatic example is the Lieb lattice-a two-dimensional geometry consisting of a square lattice with additional sites at the unit-cell edges. Owing to destructive interference, it hosts a perfectly flat band at zero energy in the absence of external perturbations, making it an ideal platform for studying exotic quantum states. Experimental realizations span ultracold atomic setups\cite{taie2015coherent, ozawa2019topological}, photonic lattices\cite{vicencio2015observation}, polariton lattices\cite{whittaker2021optical}, and superconducting circuits\cite{bello2019}.

The Lieb lattice, consisting of a square lattice with additional sites at the midpoints of each edge, is a paradigmatic platform for exploring flat-band physics due to its unique band structure, which features a perfectly dispersionless flat band intersecting dispersive bands at a Dirac-like point\cite{tang2011high, neupert2011fractional, leykam2018artificial}, produces a perfectly flat band arising from destructive interference, has emerged as a versatile platform for exploring unconventional quantum phenomena. In particular, its inherent band degeneracy and localization features make it highly sensitive to external perturbations, such as magnetic fields. The Lieb lattice is also experimentally realizable in ultracold atomic lattices\cite{taie2015coherent}, photonic waveguide arrays\cite{mukherjee2018}, Rydberg lattices\cite{Chen2024}, and programmable quantum circuits\cite{bello2019}. 
providing a versatile and controllable platform to connect theoretical predictions with practical implementations. These features make the Lieb lattice an ideal choice for studying magnetic-field-induced band deformation and tunable flat-band phenomena.

\begin{figure*}[t]
    \centering
    \includegraphics[width=0.8\textwidth]{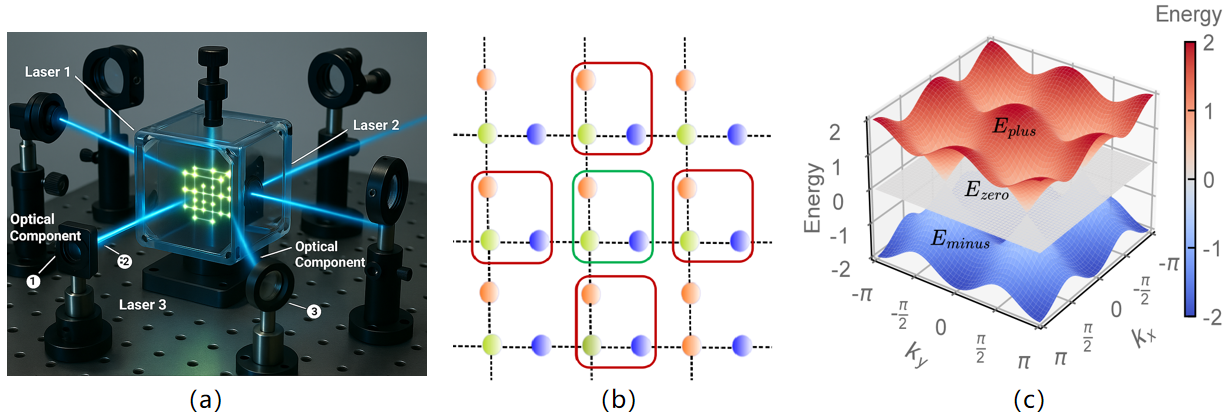}
    \caption{(Color online) Band structure for an engineered Lieb lattice. (a) Atoms(like $^{6}Li$, $^{87}Rb$) are cooled to near absolute zero using magneto-optical traps(MOTs) and transferred into the optical lattice. (b) The sketch of Lieb lattice with three sublattices in A(orange), B(yellow green) and C(purple). (c) The corresponding band structure for Lieb lattice in the absence of any external field.}
    \label{fig:lbsketchEnergy}
\end{figure*}
Similar to how thermal fluctuations modify Bloch oscillations in atomic systems\cite{yin2024influence}, magnetic fields can induce nontrivial band deformations in lattices, i.e., applying a perpendicular magnetic field to the Lieb lattice modifies its band structure through two distinct mechanisms. The first is the orbital effect, in which Peierls phases alter hopping amplitudes. At a magnetic flux of $\phi =\pi$ per plaquette, destructive interference leads to the Aharonov-Bohm(AB) caging effect\cite{Li2022, Parto2019, vidal2000, Mukherjee2020, Vidal1998, Nicolau2023, Mukherjee2018PRL, mukherjee2018, mukherjee2015observation, guzman2014experimental}, where all eigenstates become strictly localized and the spectrum collapses into dispersionless bands. 
Note that for AB caging, in specific flux configurations, quantum interference can confine particles to localized regions, resulting in completely flat bands. This “caging” phenomenon has profound implications for quantum transport, localization–delocalization transitions, and correlated many-body states.
The second is Zeeman effect\cite{roushan2017chiral, lorentz1897theory, demtroder2014atoms, zeeman1897effect, Born1999, sakurai2017modern, griffiths2017quantum, foot2005atomic, feynman1963lectures}, whereby the magnetic field couples to the spin degree of freedom, lifting spin degeneracy and inducing spin-resolved shifts in the band dispersion. Note that, Zeeman Splitting, i.e., Spin–magnetic field coupling, introduces an additional degree of band deformation, lifting degeneracies and enabling spin-resolved control of energy spectra.

In flat-band systems, the competition between AB caging and Zeeman splitting produce a rich interplay between localization, band deformation, and spin-selective transport. While each effect has been studied independently, their combined influence in the Lieb lattice remains largely unexplored. Furthermore, the combined influence of AB caging and Zeeman splitting in a Lieb lattice has not been thoroughly characterized though there exists the theoretical richness of these effects. Understanding this interplay is crucial for several reasons: it offers a clean testbed for exploring magnetic-field-driven localization and spin-resolved band engineering in a flat-band system. Mealwhile, realizations in photonic\cite{roushan2017chiral, guzman2014experimental}, cold atom\cite{shen2010single}, and electronic systems allow controllable experimental verification of magnetic field effects on flat bands. Precise control over localization and spin degrees of freedom could inform the design of spintronic devices, quantum memories, and topological qubits. By systematically investigating magnetic field–induced band deformation in the Lieb lattice, this study aims to bridge the gap between theory and experiment, providing deeper insight into correlated flat-band physics under tunable magnetic perturbations.

In this work, we present a systematic study of magnetic field induced band deformation in the Lieb lattice by simultaneously incorporating orbital Peierls phases and Zeeman spin splitting. Sec.\ref{ExperimentSetup} outlines a detailed proposal for realizing a Lieb lattice in ultracold atom systems. The evolution of the band structure in the absence of a magnetic field is analyzed in Sec.\ref{EnergyBandNoMagneticField}, whereas the corresponding behavior in the presence of a magnetic field is discussed in Sec.\ref {EnergyBandYesMagneticField}. The density of states(DOS) as a function of magnetic flux $\phi$ and Zeeman field $B_z$ is presented in Sec.\ref{DifferenceOf0andPi}, with particular emphasis on the regimes $\phi=0$ and $\phi=\pi$, where AB caging leads to pronounced flatness, and on its deformation under varying Zeeman splitting. 
The corresponding analysis of the physical discussion for magnetic field induced band deformation in a Lieb lattice is illustrated in Sec.\ref{Physical Discussion}. 

Furthermore, Sec.\ref{High-Symmetry Points} discusses the standard high-symmetry path in the two-dimensional Brillouin zone of Lieb lattice. By computing DOS evolution across Zeeman fields, we reveal how the interplay between gauge-induced localization and spin polarization drives band restructuring and energetic asymmetry, which are relevant to current efforts in synthetic quantum systems, including cold atoms in optical lattices\cite{wirth2011evidence, taie2015coherent}, topological photonic lattices\cite{guzman2014experimental,ozawa2019topological}, and superconducting circuit arrays\cite{bello2019}. Our results provide guiding principles for engineering tunable flat-band phenomena through combined magnetic control and spin manipulation. In addition, the conclusion holds in the Sec.\ref{Conclusion}.
\section{Engineering a Lieb Lattice in Ultracold Atom Systems} \label{ExperimentSetup}
\begin{figure*}[t]
    \centering
    \includegraphics[width=\textwidth]{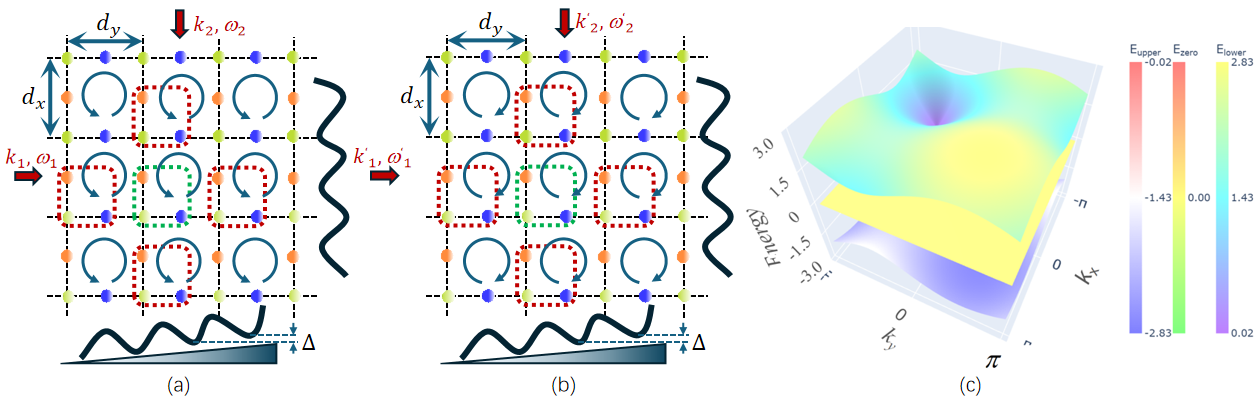}
    \caption{(Color online) Experimental schematic of a Lieb lattice implemented with ultracold atoms in an optical lattice under a synthetic magnetic field, which is consists of a 3D optical lattice, where the vertical lattice isolates different planes. The lattice constants within each plane are $|\vec{d_i}|=\lambda_i/2$, where $i=x,y$. Along $y$, bare tunneling occurs with strength J, while tunneling along $x$ is inhibited by a magnetic field gradient $B^\prime$, which introduces an energy offset between neighboring sites of (a) $\Delta$ for $|\uparrow\rangle$ atoms and (b) $\Delta$ for $|\downarrow\rangle$ atoms. Note that an additional pair of laser beams (red arrows) with wave vectors $|\vec{k}_1|\approx|\vec{k}_2|=2\pi/\lambda_K$ and frequency difference $\omega=\omega_1-\omega_2=\Delta/\hbar$ is to restore resonant tunneling with complex amplitude $K$. This realizes an effective flux of (a) $\phi=\pi/2$ for $|\uparrow\rangle$ atoms and (b) $\phi$ for $|\downarrow\rangle$ atoms. The Lieb lattice geometry features three inequivalent sites per unit cell. Complex hopping amplitudes are realized through laser-assisted tunneling, effectively generating a uniform artificial magnetic flux $\phi$ per plaquette. For comparison, there hold the corresponding primed parameters ($k_1', \omega_1'$), ($k_2', \omega_2'$). (c) Energy bands for the Lieb lattice exposed in a magnetic field.}
    \label{fig:LiebSketchM_energy_bands}
\end{figure*}

The band structure of an engineered Lieb lattice is a central topic in condensed matter and cold atom physics due to the presence of a flat band, Dirac cones, and topologically nontrivial features depending on perturbations (e.g., spin-orbit coupling, magnetic flux, interactions).
Flat bands can be engineered in cold atom systems(like $^{6}Li$, $^{87}Rb$) by designing specific photolattices(see Fig. \ref{fig:LiebSketchM_energy_bands}(a))\cite{julku2016geometric, caceres2022controlled, zeng2024transition, leung2020interaction}. Especially, 
advances in laser technology, such as the development of all-fiber systems for creating all-optical $^{87}Rb$ Bose–Einstein condensates, enhance the experimental feasibility of probing magnetic-field-induced band deformations in engineered lattices\cite{li2023all}.

Lattices are created by interfering laser beams to form periodic potentials that trap ultracold atoms. For example, Lieb and kagome lattices have been used to create flat bands with minimal dispersion\cite{taie2015coherent, taie2020spatial, jo2012ultracold, li2024engineering, nandy2025unified}. These lattices can be tuned by adjusting the laser intensity and phase, allowing for the observation of localization phenomena. Atoms (like $^{6}Li$, $^{87}Rb$) are cooled to near absolute zero using magneto-optical traps (MOTs) and transferred into the optical lattice. One can observe band occupation and momentum distribution, explore quantum dynamics, e.g., quenching or time evolution, even probe topological properties, especially when synthetic gauge fields are introduced. 

The combination produces a lattice geometry featuring three sites per unit cell (A, B, and C), characteristic of a Lieb lattice, shown the engineering process and the sketch in Figs. \ref{fig:LiebSketchM_energy_bands}(a) and (b), respectively. The superlattice term introduces additional potential minima at the plaquette centers, allowing for control of site-specific energies and tunnelings. The realization of a uniform synthetic magnetic flux and the observation of AB caging in photonic lattices have been investigated theoretically\cite{chang2021nonlinear} and experimentally\cite{PhysRevLett.121.075502, guzman2014experimental, leykam2018artificial}.
The optical potential used to realize the Lieb lattice geometry is constructed by superimposing a primary square lattice with a longer-period superlattice. The resulting potential landscape is given by
\begin{equation}
\begin{split}
    V(x, y) =& -V_s \left[ \cos^2(k x) + \cos^2(k y) \right]
    \\&- V_p \cos^2\left( \frac{k x}{2} \right) \cos^2\left( \frac{k y}{2} \right),
\end{split}
\end{equation}
where $V_s$ denotes the depth of the primary square lattice, and $V_p$  represents the strength of the superlattice modulation. The wave vector \( k=2\pi / \lambda \) corresponds to the wavelength $\lambda$ of the laser beams.

Experimental realizations of artificial magnetic fields with ultracold atoms have been achieved in optical lattices, where laser-assisted tunneling in tilted potentials generates spatially dependent complex tunneling amplitudes. In such systems, atoms accumulate phase shifts equivalent to the AB phase, enabling the direct implementation of the Hofstadter Hamiltonian and the observation of cyclotron orbits\cite{aidelsburger2013realization,miyake2013realizing}. Furthermore, using two atomic spin states with opposite magnetic moments, these setups naturally realize a time-reversal-symmetric Hamiltonian, providing a cold-atom platform for exploring the quantum spin Hall effect\cite{jotzu2014experimental,goldman2016topological, goldman2014}.
Using scanning tunneling microscopy(STM) imaging, wave-function mapping, and spectroscopy, the team directly probed the electronic structure, confirming hallmark features of the Lieb lattice: Dirac cones, flat bands, and higher-order super-Lieb patterns at elevated energies\cite{slot2017experimental,drost2017topological}. 

As a result, the experimental schematic of a Lieb lattice implemented with ultracold atoms in an Lieb optical lattice under a synthetic magnetic field, which is consists of a 3D optical lattice, is shown in Fig. \ref{fig:LiebSketchM_energy_bands}. Note that an additional pair of laser beams (red arrows) with wave vectors $|\vec{k}_1|\approx|\vec{k}_2|=2\pi/\lambda_K$ and frequency difference $\omega=\omega_1-\omega_2=\Delta/\hbar$ is to restore resonant tunneling with complex amplitude $K$. This
realizes an effective flux $\phi=\pi/2$ for $|\uparrow\rangle$ atoms and -$\phi$ for $|\downarrow\rangle$ atoms in Fig. \ref{fig:LiebSketchM_energy_bands} (a) and (b), respectively.
Lieb lattice geometry features three inequivalent sites per unit cell. Complex hopping amplitudes are realized through laser-assisted tunneling, effectively generating a uniform artificial magnetic flux $\phi$ per plaquette. The Peierls phases imprinted on the tunneling paths give rise to magnetic effects analogous to those in the Hofstadter model.
\section{Energy Band In the absence of external field} \label{EnergyBandNoMagneticField}
Lieb lattice, a two-dimensional lattice with a specific arrangement of sites(corner sites(olive green), horizontal edge sites(purple), and vertical edge sites(orange)) can be described using a tight-binding model, which is shown in Fig. \ref{fig:lbsketchEnergy}(b). The Hamiltonian for Lieb lattice in the absence of external fields is given by the tight-binding Hamiltonian in real space 
\begin{equation}
    H=-t \sum_{\langle i,j \rangle} \left( A_i^\dagger B_j + A_i^\dagger C_j + \text{h.c.} \right),
\end{equation}
where $A_i^\dagger$, $B_j^\dagger$, $C_j^\dagger$ are creation operators on sublattices $A$, $B$, and $C$. The sum $\langle i,j \rangle$ runs over nearest neighbors.

Then, the tight-binding Hamiltonian for Lieb lattice in the momentum space can be written as
\begin{equation}
    H(\mathbf{k})=\epsilon_0 \mathbf{I} + t \sum_{\delta} e^{i \mathbf{k} \cdot \delta} \mathbf{C}_{\delta},
\end{equation}
where $\epsilon_0$ is the on-site energy, $t$ is the hopping parameter, $\delta$ represents the nearest-neighbor vectors, $\mathbf{C}_{\delta}$ are the hopping matrices.
The energy dispersion relation is
\begin{equation}
 (E - \epsilon_0)^3 - t^2 e^{i k_x a} e^{-i k_x a} (E - \epsilon_0) - t^2 e^{i k_y a} e^{-i k_y a} (E - \epsilon_0)=0.   
\end{equation}
In the absence of magnetic and DC fields, the flat band in the Lieb lattice is characterized by zero dispersion. The energy of the flat band is determined by the on-site energies and the hopping parameters. The flat band is highly degenerate, leading to enhanced electron-electron interactions and the emergence of strongly correlated states\cite{wu2025flat}.
When $\epsilon_0$=0, the band structure for the Lieb lattice in the absence of any external field is as in Fig. \ref{fig:lbsketchEnergy}(c).
Due to the destructive interference of the electron wave functions at the corner and edge sites\cite{leykam2018artificial, flach2014detangling, chang2021nonlinear, mallick2022antipt, mallick2021wannierstark, yang2024realization, wu2025flat}, flat band excists, characterized by a dispersionless energy band, where the energy remains constant regardless of the momentum of the particles.
These lattices can be tuned by adjusting the laser intensity and phase, allowing for the observation of localization phenomena.
\section{Energy Band In the Presence of a Magnetic Field} \label{EnergyBandYesMagneticField}
\begin{figure*}[t]
    \includegraphics[width=0.78\textwidth]{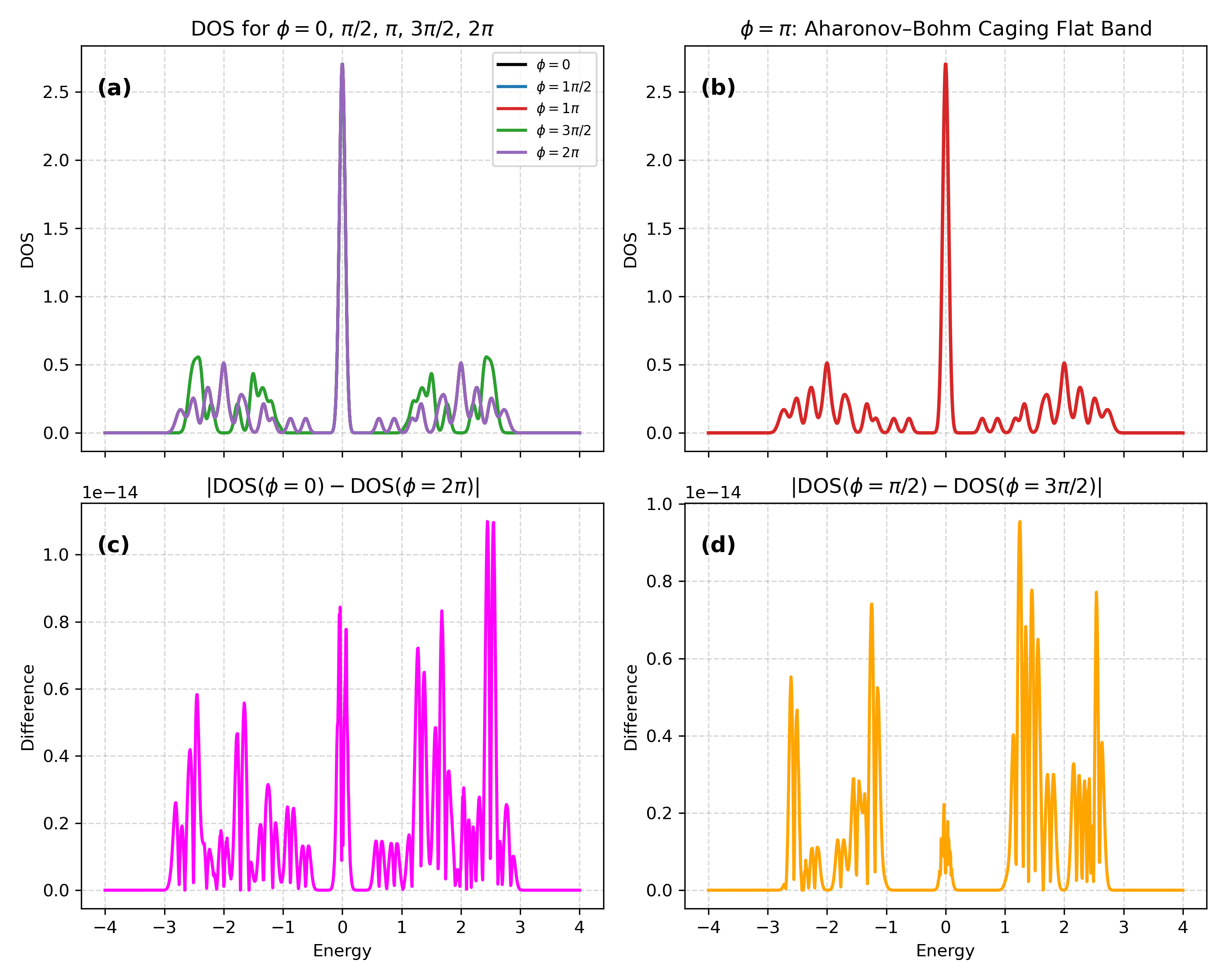}
    \caption{(Color online) (a) Density of states(DOS) of the Lieb lattice under various magnetic flux values $\phi=0, \pi/2, \pi, 3\pi/2$, and $2\pi$, showing spectral evolution and the emergence of band flattening. (b) The DOS at $\phi=\pi$, where the AB caging leads to a prominent flat band at zero energy. (c) Absolute difference between the DOS at $\phi=0$ and $2\pi$, illustrating the recovery of magnetic periodicity. (d) Absolute difference between the DOS at $\phi=\pi/2$ and $3\pi/2$, highlighting field-reversal symmetry around $\phi=\pi$.}
    \label{fig:lieb_dos}
\end{figure*} 

The tight-binding Hamiltonian for the Lieb lattice is constructed with nearest-neighbor hopping $t=1$ and periodic boundary conditions on a \(10 \times 10\) lattice. A perpendicular magnetic field introduces a Peierls phase in the hopping amplitudes, implemented as a complex exponential factor along the $y$-direction proportional to the flux $\phi$ per unit cell. The full Hamiltonian matrix of size \(300 \times 300\) is diagonalized to obtain all eigenvalues, and the DOS is computed by histogramming these eigenvalues with 200 bins and proper normalization.

The DOS is computed by evaluating the spectrum of the tight-binding Hamiltonian and applying Gaussian broadening to approximate the delta functions. Specifically, the DOS is given by
\begin{equation}
D(E)=\frac{1}{N \sqrt{2\pi \sigma^2}} \sum_n \exp\left( -\frac{(E - E_n)^2}{2\sigma^2} \right),
\end{equation}
where $E_n$ are the eigenvalues of the Hamiltonian, $N$ is the total number of eigenstates, and $\sigma$ is the broadening parameter. In our calculations, we choose $\sigma=0.05$, which provides a smooth approximation while preserving fine spectral features.

Fig. \ref{fig:lieb_dos} illustrates the evolution of the DOS in Lieb lattice as a function of magnetic flux $\phi$. Panel (a) shows the DOS for $\phi=0, \pi/2, \pi, 3\pi/2$, and $2\pi$. At zero flux, the system exhibits a characteristic flat band at $E=0$ and two dispersive bands. As the flux increases, spectral gaps open and the band structures are progressively deformed. 
Panel (b) presents the DOS at $\phi=\pi$, where the effect of AB caging becomes pronounced. The destructive interference induced by the $\pi$-flux leads to complete localization of wavefunctions within confined plaquette structures, giving rise to a macroscopic flat band centered at $E=0$. This is a hallmark signature of AB caging, previously predicted for line-centered lattices under half-integer flux quanta.
To verify the magnetic periodicity of the system, panel (c) shows the absolute difference between the DOS at $\phi=0$ and $2\pi$, which vanishes up to numerical precision. This confirms that the spectrum is $2\pi$-periodic with respect to the magnetic flux, as expected from the Hofstadter framework. 
Panel (d) compares the DOS between $\phi=\pi/2$ and $3\pi/2$, revealing a near-symmetric structure. This reflects a field-reversal symmetry about $\phi=\pi$, indicating that the system respects time-reversal invariance under a flux inversion about the half-flux point. Such symmetry analysis reinforces the topological and interference-driven nature of the band evolution in the Lieb lattice under magnetic fields.
\begin{figure*}[t]
    \centering
    \includegraphics[width=0.7\linewidth]{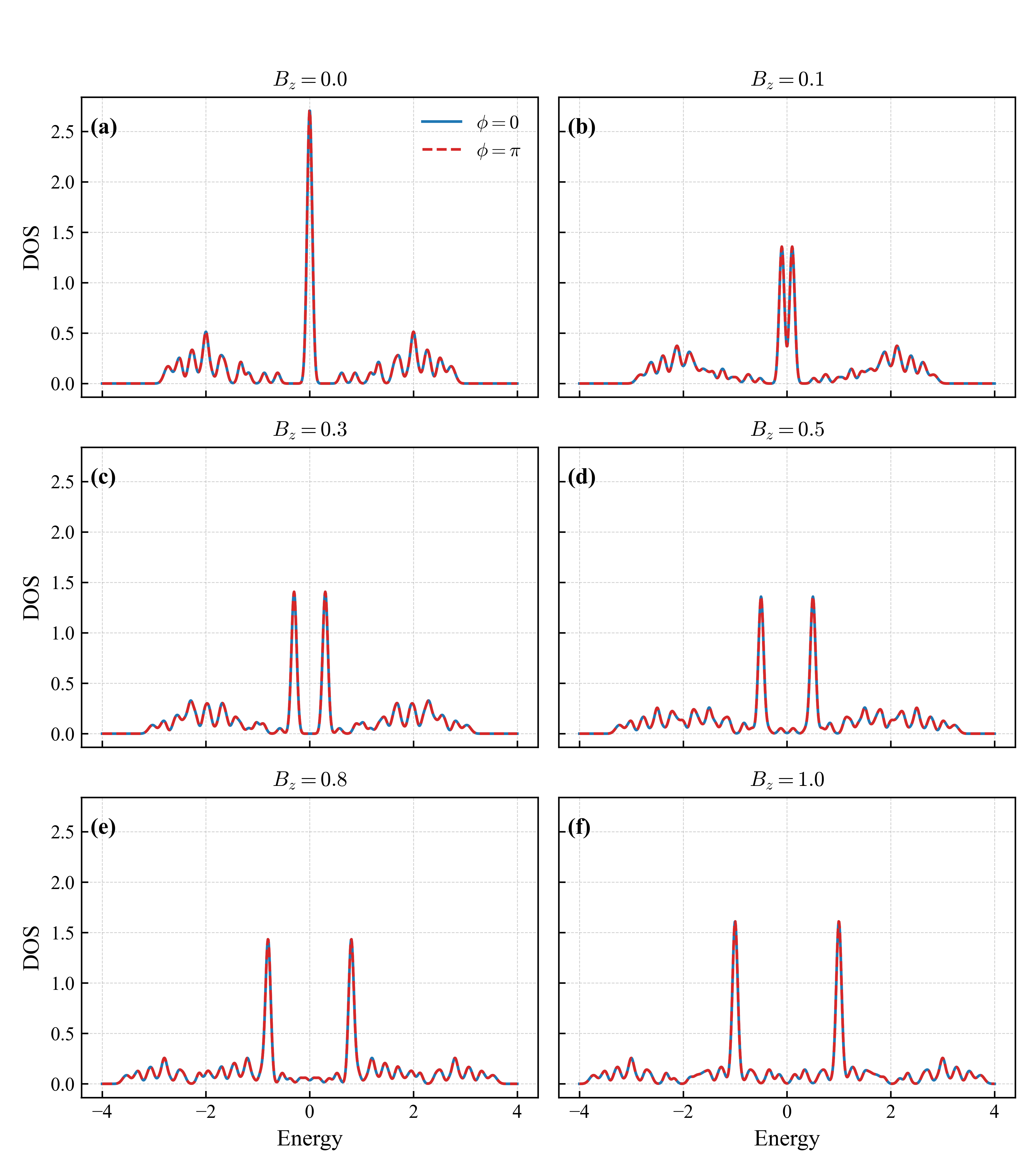}
    \caption{(Color online) The DOS of Lieb lattice under varying Zeeman splittings $B_z=0$, $0.1$, $0.3$, $0.5$, $0.8$, and $1.0$ for two representative magnetic flux values: $\phi=0$(solid blue) and $\phi=\pi$(dashed red). Each panel (a)-(f) corresponds to a specific $B_z$, showing how the spin-resolved energy levels evolve. At $\phi =\pi$, the AB caging effect generates a nearly flat band at zero energy, which becomes Zeeman-split and eventually broadened as $B_z$ increases. In contrast, the $\phi =0$ case lacks such flatness and exhibits smoother shifts of spectral weight.The calculations use a $10 \times 10$ Lieb lattice with Gaussian broadening $\sigma=0.05$.}
\label{fig:DOS_Zeeman_phi0_phiPi}
\end{figure*}

We consider a tight-binding Hamiltonian on a two-dimensional Lieb lattice under the influence of both a uniform magnetic field and a Zeeman field. The system contains three sublattices per unit cell and is defined on a $L_x \times L_y$ square grid with periodic boundary conditions. The magnetic field is introduced via Peierls substitution, which assigns complex hopping phases according to the vector potential. For a magnetic flux $\phi$ per plaquette, the hopping along the $A$-$C$ bonds acquires position-dependent Peierls phases $e^{\pm i 2\pi \phi x / 2\pi}$, while the $A$–$B$ bonds remain unmodified.
Including Zeeman effect as a spin-dependent onsite energy shift, the total Hamiltonian becomes spin-block-diagonal, i.e., 
\begin{equation}
H_{\text{total}}=
\begin{pmatrix}
H_0 + B_z \mathbb{I} & 0 \\
0 & H_0 - B_z \mathbb{I}
\end{pmatrix},
\end{equation}
where $H_0$ is the spinless tight-binding Hamiltonian with magnetic flux, and $B_z$ is the out-of-plane Zeeman field strength.

The density of state(DOS) is computed by diagonalizing $H_{\text{total}}$ and applying Gaussian broadening
\begin{equation}
D(E)=\frac{1}{\sqrt{2\pi} \sigma} \sum_n \exp\left[-\frac{(E - E_n)^2}{2\sigma^2}\right],
\end{equation}
with $\sigma=0.05$ and energy sampled over a grid $E \in [-4, 4]$.
\section{Zeeman-induced Spin Splitting of DOS at $\phi=0$ and $\phi=\pi$}\label{DifferenceOf0andPi}
Fig. \ref{fig:DOS_Zeeman_phi0_phiPi} presents the evolution of DOS under increasing Zeeman field $B_z$, for two flux values $\phi=0$ and $\phi=\pi$.
At zero flux ($\phi=0$), the spectrum exhibits a dispersive structure with a broad DOS distribution. As $B_z$ increases, spin degeneracy is lifted, and the DOS gradually separates into two components shifted symmetrically about zero energy, reflecting the spin-up and spin-down channels. In contrast, at $\phi=\pi$, the system exhibits a prominent flat band at $E=0$, which originates from the destructive interference and localization known as the AB caging effect. This flat band gives rise to a sharp DOS peak. Upon introducing $B_z$, the degeneracy is lifted, and the single peak splits into two narrow ones, symmetrically displaced. At higher $B_z$, these peaks broaden and eventually merge with the continuum bands.

The stark difference between the $\phi=0$ and $\phi=\pi$ cases highlights the role of magnetic flux in shaping the spectral flatness and its sensitivity to Zeeman perturbations. The $\phi=\pi$ flat band is particularly susceptible to spin splitting due to its highly localized nature, while the $\phi=0$ states respond more smoothly due to their dispersive character. These features are of interest for cold atom emulations of flat-band systems, as well as for exploring interaction effects in spin-split flat-band platforms.
\section{Physical Discussion} \label{Physical Discussion}
The application of a perpendicular magnetic flux $\phi$ threading each unit cell modifies the band structure through complex hopping phases introduced by the Peierls substitution. This results in the splitting of the energy bands into magnetic subbands, which can exhibit fractal-like patterns (e.g., Hofstadter butterfly) at rational values of the flux.

In Lieb lattice, the presence of magnetic field notably affects the flat band by inducing band splitting and opening of gaps. The interference between hopping paths due to the magnetic field alters localization properties and modifies the DOS. At $\phi=\pi$, these effects are especially pronounced, causing significant rearrangement of the energy levels and the formation of distinctive spectral features arising from the lattice symmetry and magnetic flux.
\subsection{Aharonov-Bohm Effect}
When a magnetic field is applied perpendicular to the Lieb lattice, the Hamiltonian is modified to include the vector potential $\mathbf{A}$
\begin{equation}
H_B=-t \sum_{\langle i,j \rangle} e^{i \frac{e}{\hbar c} \int_{i}^{j} \mathbf{A} \cdot d\mathbf{r}} (c_i^\dagger c_j + \text{h.c.}) + \sum_i \epsilon_i c_i^\dagger c_i.
\end{equation}
AB effect($\phi=\pi$) profoundly influences electron motion in a Lieb lattice by inducing phase shifts that lead to destructive interference and electron localization\cite{zohuri2024aharonov}.This effect, known as AB caging, results in perfectly flat bands in the energy spectrum. However, the presence of electron-electron interactions can disrupt this caging effect, leading to delocalization and changes in the electronic properties of lattice.

Thus, the Hamiltonian for the Lieb lattice with AB cage effect can be written as
\begin{equation}
\begin{split}
H &= \sum_{m,n}\Big(t_{AB}(e^{i\theta_{m,n}}c^\dagger_{A_{m,n}}c_{B_{m,n}}+\text{h.c.}) \\
&+ t_{AC}( c^\dagger_{A_{m,n}} c_{C_{m,n}} + \text{h.c.})
+ t_{BC}(c^\dagger_{B_{m,n}} c_{C_{m,n}} + \text{h.c.})\Big),
\end{split}
\end{equation}
where $t_{AB}$, $t_{AC}$, and $t_{BC}$ are the hopping parameters between different sites, $c^\dagger$ and $c$ are the creation and annihilation operators, respectively, and $\theta_{m,n}=\alpha m$ represents the phase shift due to the magnetic field, with $\alpha$ being the magnetic flux through the unit cell. The Magnetic Field Strength, i.e., magnetic flux $\phi$ through each unit cell is given by $\phi=\alpha \phi_0$, where $\phi_0=h/e$ is the flux quantum. For the AB cage effect to be significant, $\alpha$ is typically chosen to be $1/2$, corresponding to a flux of $\phi=\phi_0/2$.

To calculate the energy bands, we consider the tight-binding model in the momentum space. The Bloch Hamiltonian for the Lieb lattice with the AB cage effect can be written as
\begin{equation}
\begin{split}
H(\mathbf{k}) &=t_{AB} \left( e^{i(k_x + \alpha)} \sigma_x + e^{-i(k_x + \alpha)} \sigma_x \right) \\
& + t_{AC} \left( e^{ik_y} \sigma_y + e^{-ik_y} \sigma_y \right)  \\
& + t_{BC} \left( e^{ik_x} \sigma_z + e^{-ik_x} \sigma_z \right),
\end{split}
\end{equation}
where $(\mathbf{k}=(k_x, k_y))$ is the wave vector, and $(\sigma_x)$, $(\sigma_y)$, $(\sigma_z)$ are the Pauli matrices.
The energy bands are obtained by solving the eigenvalue problem $H(\mathbf{k}) \psi=E(\mathbf{k}) \psi$. For $\alpha=1/2$(The ratio of the magnetic flux in each unit cell to 2$\pi$), the energy bands exhibit significant changes due to the AB cage effect. The flat band becomes completely flat, and the other bands show non-trivial dispersion. The energy bands can be plotted as a function of the wave vector $(\mathbf{k})$.

Thus, AB cage effect significantly alters the energy band structure of the Lieb lattice, leading to the complete flattening of the flat band and non-trivial dispersion of other bands, which introduces a phase shift in the electron wave functions, altering the interference patterns and, consequently, the energy levels of the flat band. In the Lieb lattice, the AB effect can lead to a dispersion of the flat band, meaning that the originally flat band can acquire a non-zero bandwidth. This dispersion arises because the phase shift depends on the electron's path and the magnetic flux through the loops formed by the lattice sites. The energy levels of the flat band can change and spread out, leading to a more complex band structure. This effect provides a unique platform for studying the interplay between magnetic fields and electronic states in two-dimensional lattices.
\subsection{Zeeman Splitting}
\begin{figure}[htbp]
    \centering
    \includegraphics[width=\linewidth]{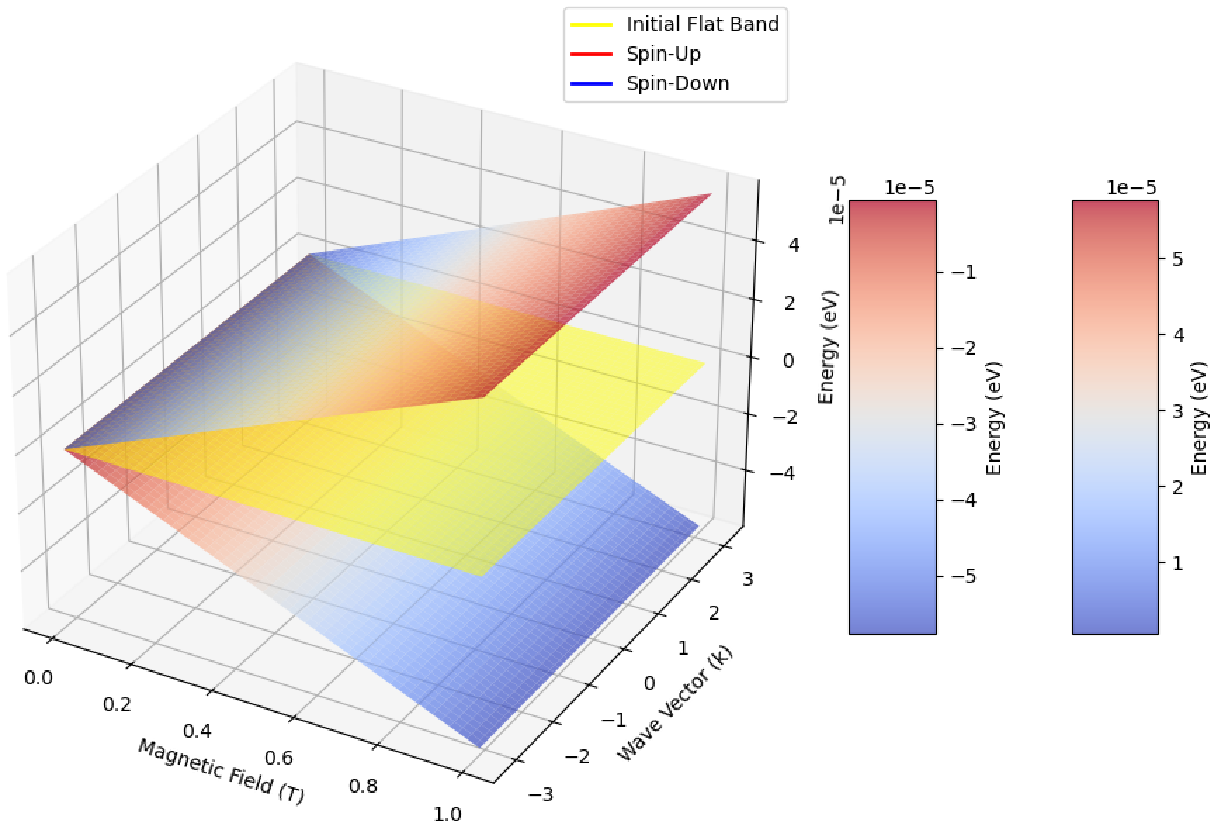}
    \caption{(Color online) The influence of Zeeman effect on the energy structure fo a Lieb lattice in the presence of magnetic field, i.e., the flat band of the energy levels split into two branches corresponding to spin-up $E_{spin-up}$ and spin-down $E_{spin-down}$ electrons, respectively.}
    \label{fig:Zeeman effect}
\end{figure}
The application of a magnetic field to the Lieb lattice can significantly affect the energy levels of the flat band through several key mechanisms: the Zeeman effect, AB effect, and modifications to the band structure. Zeeman effect arises because electrons have an intrinsic magnetic moment due to their spin. When a magnetic field is applied, this magnetic moment interacts with the field, causing the energy levels to split. This splitting is proportional to the strength of the magnetic field and the spin orientation of the electrons.This splitting lifts the degeneracy of the flat band and affects the overall energy structure.
\begin{figure*}[t]
  \centering
  \includegraphics[width=0.9\linewidth]{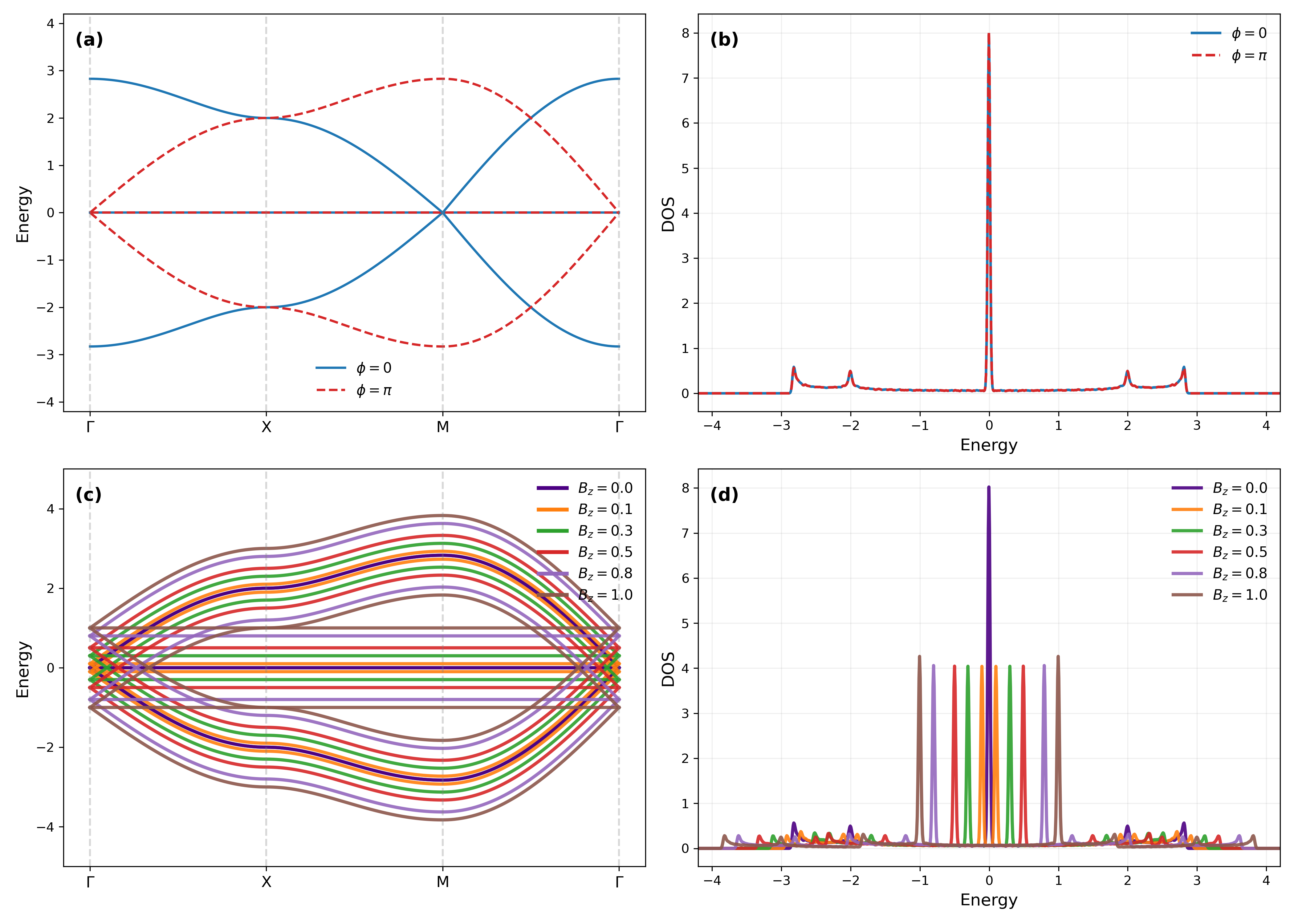}
  \caption{(Color online) (a) Band structures of the Lieb lattice along the high-symmetry path $\Gamma \to X \to M \to \Gamma$ for magnetic flux $\phi=0$ (solid blue lines) and $\phi=\pi$ (dashed red lines). The flat bands at $\phi=\pi$ result from the AB caging effect. (b) Corresponding DOS for $\phi=0$ and $\phi=\pi$, illustrating the emergence of flat bands as sharp peaks in the DOS at $\phi=\pi$. (c) Band structures at $\phi=\pi$ with Zeeman splitting induced by various $B_z$ fields (colors indicate different Zeeman strengths), showing spin-resolved splitting of the previously flat bands into six spin-polarized bands. (d) Corresponding DOS with Zeeman splitting, revealing the gradual splitting and broadening of flat-band peaks with increasing $B_z$. The high-symmetry points in the Brillouin zone are $\Gamma=(0,0)$, $X=(\pi,0)$ and $M=(\pi,\pi)$.}
    \label{fig:lieb_band_dos}
\end{figure*}
\begin{equation}
H_Z=\frac{1}{2} g \mu_B B \sum_i \sigma_z c_i^\dagger c_i,
\end{equation}
where $g$ is the Landé $g$-factor (typically $g \approx 2$ for electrons),
$\mu_B$ is the Bohr magneton $\mu_B=\frac{e \hbar}{2m_e}$, $B$ is the magnetic field strength, $\sigma_z$ is the Pauli matrix for spin, $c_i^\dagger$ and $c_i$ are the creation and annihilation operators for electrons at site $i$. Zeeman effect increases the spin degeneracy of the flat band. In the absence of a magnetic field, the flat band is typically spin-degenerate (i.e., electrons with spin up and spin down have the same energy). With a magnetic field, the energy levels split into two branches corresponding to spin-up and spin-down electrons(see Fig. \ref{fig:Zeeman effect}). This splitting is linear with the magnetic field strength
\begin{equation}
\begin{split}
    E_\text{spin-up} &= E_0 + \frac{1}{2} g \mu_B B, \\
    E_\text{spin-down} &= E_0 - \frac{1}{2} g \mu_B B,
\end{split}
\end{equation}
where $E_0$ is the energy of the flat band in the absence of the magnetic field.
\section{High-Symmetry Points in the Brillouin Zone} \label{High-Symmetry Points}
The band structures are calculated along a standard high-symmetry path in the two-dimensional Brillouin zone of the Lieb lattice. The chosen path is
$\Gamma \to X \to M \to \Gamma$, where the high-symmetry points correspond to $\Gamma=(0,0), X=(\pi, 0), M=(\pi, \pi)$. These points represent special locations in the reciprocal space that reflect the underlying lattice symmetries and boundary conditions. Calculating the energy bands along this path captures the essential features of the dispersion relations and allows one to identify critical phenomena such as band crossings, gaps, and flat bands. Therefore, this choice of high-symmetry path enables clear and physically meaningful comparisons between different scenarios, such as zero magnetic flux, $\phi=\pi$ flux exhibiting AB caging, and various Zeeman field strengths inducing spin splitting.

The band structures shown in Fig. \ref{fig:lieb_band_dos}(a) and (c) are plotted along the high-symmetry path $\Gamma \to X \to M \to \Gamma$ in the two-dimensional Brillouin zone of the Lieb lattice, where $\Gamma=(0,0)$, $X=(\pi,0)$, and $M=(\pi,\pi)$. This path captures the essential features of the lattice’s band topology and allows clear visualization of magnetic flux and Zeeman-induced effects.

Fig. \ref{fig:lieb_band_dos}(a) compares the band structures without magnetic flux $\phi=0$ and with maximal flux $\phi=\pi$. The application of $\phi=\pi$ induces the AB caging effect, leading to perfectly flat energy bands across the Brillouin zone. These flat bands indicate complete localization of wavefunctions due to destructive interference caused by the magnetic phases. The corresponding DOS in Fig. \ref{fig:lieb_band_dos}(b) reflects these changes: the flat bands at $\phi=\pi$ produce sharp, delta-function-like peaks, contrasting with the broader DOS at $\phi=0$. Fig. \ref{fig:lieb_band_dos}(c) and (d) extend the analysis by including Zeeman splitting under various out-of-plane magnetic fields $B_z$. The spin degeneracy is lifted, doubling the number of bands to six, with clear splitting proportional to $B_z$. This spin splitting modifies both the band dispersion and the DOS, which gradually evolves from sharp single peaks into multiple spin-polarized peaks, indicating spin-resolved localized states.

Fig. \ref{fig:lieb_band_dos}(a) and (b) show the flattening of the energy bands and the emergence of a prominent zero-energy peak in the DOS at $\phi=\pi$, reflecting the AB caging effect and the corresponding localization of electronic states. In contrast, Figs. \ref{fig:lieb_band_dos}(c) and (d) reveal the spin-dependent deformation of bands and splitting of the flat band into spin-polarized subbands as the Zeeman energy $B_z$ increases. This splitting leads to additional fine structure in the DOS, with clear signatures of band separation and spin-resolved flattening.

Lieb lattice provides a paradigmatic platform to study the interplay of magnetic field effects on band topology and localization phenomena. Experimental advances in controlling excited-band lifetimes via lattice-depth optimization\cite{shui2023optimal} highlight the practicality of realizing and probing band deformation effects in optical Lieb lattices. The introduction of a uniform magnetic flux $\phi$ through the lattice plaquettes induces Peierls phases, modifying the hopping amplitudes and leading to the celebrated AB caging effect at $\phi=\pi$. This effect manifests as complete flattening of energy bands, reflecting perfectly localized eigenstates that cannot propagate due to destructive interference of electron wavefunctions.

Consequently, the combination of Zeeman splitting and AB-induced dispersion can lead to a rich variety of energy level structures. The flat band can split into multiple sub-bands with different dispersions. The energy levels of the flat band can shift upwards or downwards depending on the spin orientation and the specific path taken by the electrons in the lattice. In some cases, the magnetic field can induce transitions between different quantum states, leading to the emergence of new phases or states of matter (e.g., fractional quantum Hall states).
These effects highlight the interplay between quantum mechanics and external fields, and they provide a rich playground for exploring novel quantum phenomena in condensed matter systems.
\section{Conclusion} \label{Conclusion}
In summary, we investigate the magnetic-field-induced band deformation in a Lieb lattice by incorporating both Peierls substitution and Zeeman splitting. Our calculations reveal that AB caging and Zeeman splitting act in concert to reshape flat bands, localization, and spin textures. These results demonstrate that magnetic flux and spin splitting provide a powerful means of tuning band properties in synthetic lattices. With recent progress in cold-atom quantum simulators and optical lattices, the Lieb lattice emerges as a natural platform for probing magnetic-field-driven band deformation and engineering correlated and topological phases. 

Our results highlight that the interplay between magnetic flux and Zeeman splitting provides a powerful tuning mechanism for manipulating flat bands, localization, and spin textures in synthetic quantum lattices. These findings are directly relevant to ongoing experimental efforts in engineered cold-atom systems, photonic lattices, and programmable quantum simulators, where control over magnetic responses is essential for realizing correlated and topological phases.The tunability of ultracold atoms, including the observation of scattering halos in Feshbach molecular condensates\cite{chen2025scattering}, highlights the feasibility of probing magnetic field-induced band deformations in engineered Lieb lattices. The observation of collisional scattering in strongly interacting Feshbach molecules\cite{PhysRevResearch.7.023030} further underscores the potential of ultracold atoms as a platform to study magnetic-field-induced band deformation in the Lieb lattice. 
The observation of vestigial order melting in chiral atomic superfluids within optical lattices\cite{yu2025vestigial} highlights the ability of quantum simulators to probe ordered phases and band deformations in engineered lattice geometries.
The ability of atomic quantum simulators to probe dimensional crossover physics\cite{tian2025probinguniversalphasediagram} demonstrates the feasibility of realizing and exploring magnetic-field-induced band deformations in engineered Lieb lattices.
\section*{Acknowledgement}
This work is supported by the National Natural Science
Foundation of China (Grants No. 92365208) and National Key Research and Development Program of China (Grants No. 2021YFA0718300 and No. 2021YFA1400900).
\bibliographystyle{apsrev4-2} 
\bibliography{LiebRefs}
\end{document}